\documentclass[12pt,preprint]{aastex}
\shorttitle{High Velocity H~{\footnotesize I} Absorption Toward NGC~1275}
\shortauthors{Momjian et al.}

\begin{document}

%% LaTeX will automatically break titles if they run longer than
%% one line. However, you may use \\ to force a line break if
%% you desire.

\title{GLOBAL VLBI OBSERVATIONS OF THE HIGH VELOCITY \\H~{\footnotesize I}
ABSORPTION TOWARD NGC~1275}

%% Use \author, \affil, and the \and command to format
%% author and affiliation information.
%% Note that \email has replaced the old \authoremail command
%% from AASTeX v4.0. You can use \email to mark an email address
%% anywhere in the paper, not just in the front matter.
%% As in the title, you can use \\ to force line breaks.

\author{E. Momjian\altaffilmark{1, 2}}
\email{momjian@pa.uky.edu}

\author{J. D. Romney\altaffilmark{1}}
\email{jromney@nrao.edu}

\and

\author{T. H. Troland\altaffilmark{2}}
\email{troland@pa.uky.edu}

%% Notice that each of these authors has alternate affiliations, which
%% are identified by the \altaffilmark after each name.  Specify alternate
%% affiliation information with \altaffiltext, with one command per each
%% affiliation.

\altaffiltext{1}{National Radio Astronomy  Observatory, P O Box O, Socorro, NM 87801.}
\altaffiltext{2}{University of Kentucky, Department of Physics and Astronomy, Lexington, KY 40506.}

%% Mark off your abstract in the ``abstract'' environment. In the manuscript
%% style, abstract will output a Received/Accepted line after the
%% title and affiliation information. No date will appear since the author
%% does not have this information. The dates will be filled in by the
%% editorial office after submission.

\begin{abstract}
We present global VLBI observations of the $\sim$8100~km~s$^{-1}$
H~{\footnotesize I} absorption feature detected toward the strong radio nucleus
3C~84 of NGC~1275 in the Perseus cluster (${V_{\rm sys}\sim}$5200~km~s$^{-1}$).
The observations were obtained using the Very Long Baseline Array (VLBA), the
phased Very Large Array (VLA), and three stations of the European VLBI Network
(EVN)\null. Our results provide the first high dynamic range  image of this
feature at high spectral and spatial resolutions. We detect six distinct
absorption peaks with optical depths ranging from 0.1 to 0.45, and multiple weak
features with $\tau \leq$ 0.1. The compactness  of the background radio source
3C~84, which has a linear extent of only about 16~pc, limits the
conclusions that can be drawn as to the nature of the intervening object,
which must be falling toward the center of the
Perseus cluster at $\sim$3000~km~s$^{-1}$. However, the detected absorption
peaks indicate the existence of several H~{\footnotesize I} clouds with
velocity differences and widths similar to those seen in Galactic neutral
hydrogen absorption and similar to some of the H~{\footnotesize I} absorption
seen in the Large Magellanic Cloud.
The most prominent H~{\footnotesize I} clouds extend from 12 to 30~mas
(milliarcseconds) on the plane of the sky. The derived H~{\footnotesize I}
column densities, assuming $T_{\rm s}=50$~K, range over $(0.35-3.8)\times
10^{20}$~cm$^{-2}$, and the implied volume densities range between 1.4 and 11~
cm$^{-3}$. \end{abstract}

%% Keywords should appear after the \end{abstract} command. The uncommented
%% example has been keyed in ApJ style. See the instructions to authors
%% for the journal to which you are submitting your paper to determine
%% what keyword punctuation is appropriate.

\subjectheadings{galaxies: active --- galaxies: individual (NGC~1275) ---
galaxies: ISM --- radio continuum: galaxies --- techniques: interferometric}

%% From the front matter, we move on to the body of the paper.
%% In the first two sections, notice the use of the natbib \citep
%% and \citet commands to identify citations.  The citations are
%% tied to the reference list via symbolic KEYs. The KEY corresponds
%% to the KEY in the \bibitem in the reference list below. We have
%% chosen the first three characters of the first author's name plus
%% the last two numeral of the year of publication as our KEY for
%% each reference.

\section{INTRODUCTION}

NGC~1275 is a giant cD elliptical galaxy located near the center of the
Perseus cluster, with a systemic velocity of $\sim$5200~km~s$^{-1}$. The galaxy
has an active nucleus whose presence is revealed by the powerful compact radio
source 3C~84. Lying in front of NGC~1275 is another velocity system at
$\sim$8200~km~s$^{-1}$. This system is visible in H$\alpha$ emission \citep
{CAU92} as well as in 21~cm H~{\footnotesize I} absorption \citep {DEY73}. The
nature of the foreground system is unclear. It may be a late-type Sc or Sd
galaxy whose H~{\footnotesize II} regions account for the H$\alpha$ emission.
However, no nucleus or other systematic morphological structure is apparent
\citep {UNG90}. Also, it is not known if the foreground system is in a chance
alignment with NGC~1275 \citep {RUB77,RUB78} or if it is strongly interacting
with the elliptical galaxy \citep {UNG90}.

The H~{\footnotesize I} absorption near 8116~km~s$^{-1}$ consists of several
relatively narrow components (1.5--5~km~s$^{-1}$), similar in nature
to Galactic H~{\footnotesize I} absorption lines. (All velocities
in this paper are heliocentric, using the optical convention for redshift.)
The presence of such lines
offers the opportunity to study spatial structure of the absorbing
H~{\footnotesize I} in the high-velocity system. However, radio continuum
observations reported by \citet {SIJ93} indicate that 76\% of the 1395~MHz
continuum flux arises in a milliarcsecond-scale core. Therefore, very long
baseline interferometry (VLBI) techniques are essential. The first VLBI
observations on this feature were performed by \citet {ROM78}, who reported
one strong absorption peak at 8114~km~s$^{-1}$ located between two wide
absorption shoulders, with some indication of weaker absorption features. Only
the strong peak was subject to detailed study, due to the limitations of the
three element VLBI array and the equipment then available.

The completion of the Very Long Baseline Array (VLBA) in the early 1990s
provided a new opportunity to study this H~{\footnotesize I} absorption
system. An early observation of this feature with the ten stations of the
VLBA was performed in 1995 as part of a wider project, which also included the
Galactic and the low-velocity H~{\footnotesize I} absorption toward 3C~84
(J.~E.~Conway, J.~D.~Romney, M.~Rupen, \& A.~J.~Beasley, unpublished).
Analysis of the high-velocity H~{\footnotesize I}
absorption feature by EM and JDR produced interesting preliminary results. The
poor quality of the original data, and the limited observing time, led us to
propose new observations to study this absorption feature with a more sensitive
VLBI array.

In this paper, we report a detailed study of the high-velocity H
{\footnotesize I} absorption feature with a global VLBI
array. Our results show the existence of several neutral hydrogen clouds in the
intervening object, which is falling toward NGC~1275 at $\sim$3000~km~s$^{-1}$.
We adopt a distance of 104~Mpc for the Perseus cluster, assuming
${H_\circ=50}$~km~s$^{-1}$~Mpc$^{-1}$. At this distance 1~mas corresponds to
0.49~pc.

\section{OBSERVATIONS AND DATA REDUCTION}

The observations were carried out at 1383~MHz on Feb.~19, 2000
using the VLBA\footnote{The National
Radio Astronomy Observatory is a facility of the National Science Foundation
operated under cooperative agreement by Associated Universities, Inc.}, the
VLA as a phased array, and four stations of the European VLBI Network, namely
Effelsberg, Jodrell Bank (Lovell), Medicina and the Westerbork phased array.
Two different baseband channel widths were observed, wide (8~MHz) and narrow
(0.5~MHz), both with right and left-hand circular polarizations. Both bands were
centered at the frequency of the neutral hydrogen 21~cm line, at a heliocentric
velocity of $V=8114$~km~s$^{-1}$, and recorded using 2-bit sampling. The total
observing time was 25 hours, with the first 11 hours being observed only by the
EVN stations. The wide and the narrow-band data were correlated in separate
passes at the VLBA correlator in Socorro, NM with 4~seconds correlator
integration time, to produce 16 and 256 spectral channels, respectively. Table~1
summarizes the parameters of these observations.

Along with 3C~84, the radio sources J0303+4716 and J0313+4120 were
observed at the same frequencies for calibration purposes. Due to the lower
flux densities of these two nearby calibrators compared to 3C~84, which is one
of the strongest sources in the radio sky, calibration measurements were also
obtained by shifting the narrow-band frequency of 3C~84 occasionally, to an
offset frequency 500~kHz (one bandwidth) higher, away from the H~{\footnotesize
I} absorption line. Both wide and narrow-band data were reduced and processed
using the AIPS (Astronomical Image Processing System) package of NRAO.

Initial examination of the data showed the existence of interference in a
wide frequency range at the EVN station Medicina at all times.
All data from this station had to be abandoned. After many iterations,
we found that amplitude self-calibration was essential in achieving high
dynamic range images, and thus we also had to abandon the observations while
the remaining three EVN stations were observing alone.

Wide-band channels were included in the observations to provide a high
dynamic range continuum image for the background source 3C~84.
After applying {\it a priori} flagging, and manually
excising integrations affected by interference, we performed amplitude
calibration using the measurements of the antenna gain and the system
temperature ($T_{\rm sys}$) of each station. Bandpass calibration was
performed, and the spectral channels were averaged, self-calibrated, and imaged.
The resulting continuum image has a dynamic range of 17,000.

The 256 spectral channels of the narrow-band data set were used to study the
high-velocity H~{\footnotesize I} absorption feature at high spectral
resolution, with a channel separation of 0.43~km~s$^{-1}$ (1.95~kHz).
This data set was reduced twice. Each reduction implemented a
different amplitude calibration method for the purposes of comparison, and
to obtain the best possible results from our observations.
In the first reduction, as for the wide band, the amplitude calibration was
based on the measured antenna gains and $T_{\rm sys}$ values.
In the second reduction, the ``template'' spectrum method was utilized,
obtaining amplitude solutions by fitting a long-term averaged total-power
spectrum of a specific antenna to short-term (20~min) intervals on the other
stations, and applying the solutions to the cross-power spectra. This method was
originally developed for the calibration of VLBI spectral line observations
using poorly calibrated antennas. The offset-frequency 3C~84 observations were
used to correct the bandpass variations of the cross-power spectra in both data
reduction methods. The source J0303+4716 was used for bandpass calibration of
the total-power template spectrum.

The results reported in this paper are based on the narrow-band data reduction
that utilized the measured antenna gains and $T_{\rm sys}$ values for amplitude
calibration. Consistent results were obtained for the template spectrum
method, but it also introduced spectrally-dependent image artifacts which could
not be corrected by the self-calibration process. The template method may
provide superior results, however, when applied on stronger absorption or
emission-line features.

The phase and amplitude corrections from the self-calibration of the wide band
were applied on the narrow-band data sets. H~{\footnotesize I}
data cubes were constructed by subtracting the continuum in the UV-plane
using the AIPS task `UVLSF'. Optical depth cubes were calculated from the
continuum image and the inner 140 channels of each line cube as
$\tau(\nu) = - {\rm ln} [ 1 - I_{\rm line}(\nu)/I_{\rm continuum}]$.

\section{RESULTS}

\subsection{THE CONTINUUM}

Figure~1 is our continuum image of 3C~84 at 1383~MHz. The image was
reconstructed with a grid weighting intermediate between natural and uniform
(${\rm ROBUST}= 0$ in AIPS task `IMAGR').

The source
consists of a compact core and a jet extending to the south. An image of 3C~84
convolved to a larger beam size revealed the northern counterjet, which was
first discovered at 8.4~GHz \citep {WRB94} and 22~GHz \citep {VRB94},
simultaneously. This northern feature was not detected at 1.7~GHz in a high
dynamic range image by \citet {BBD}; however, \citet {STV98} detected it at
1414~MHz. The extent and the flux density of the northern counterjet in our
observations are consistent with the results of \citet {STV98}.

The total flux density of 3C~84 in our observations is $\sim$18~Jy, almost 4~Jy
less than the total flux density reported by \citet {TV96} at 1347~MHz and
\citet {STV98} at 1414~MHz. Both these VLBA observations were carried out in
1995. A similar proportional decline in the flux density of this source is
reported in the single dish monitoring observations between 1995 and 2000 at
higher frequencies (H.~D.~Aller, M.~F.~Aller, G.~E.~Latimer, \& P.~A.~Hughes, in
preparation). The decline of the flux density of 3C~84 at the frequencies
monitored has been continuous since 1982--1984.

\subsection{THE H~{\footnotesize I} ABSORPTION}

Our study reveals the existence of several H~{\footnotesize I} absorption peaks
in the high-velocity system associated with NGC~1275.
Figure~2 shows a
continuum image of 3C~84 and Hanning smoothed spectra of H~{\footnotesize I}
optical depth at various locations against the continuum image.
Six main absorption peaks can be distinguished in the
spectra of Figure~2, with velocity widths of 1.5--5~km~s$^{-1}$ at half
maximum. Weaker H~{\footnotesize I} components are also evident in the velocity
range 8128.77--8136.17~km~s$^{-1}$ .

Figure~3 presents optical-depth images covering
the velocity range 8127.0--8106.2~km~s$^{-1}$, in every other spectral
channel. Because the signal-to-noise ratio in the optical-depth images is
poor where the  continuum is weak, these images are blanked in areas
where the flux density in the background continuum image is below 1.2\% of the
peak value, i.e., less than 64~mJy~beam$^{-1}$. These images explicitly show the
variation of the H~{\footnotesize I} distribution in the small section of the
foreground galaxy which is seen against the bright compact background source.
The shift in the neutral hydrogen opacity from west to east is apparent in
the images that cover the velocity range 8119.2--8113.1~km~s$^{-1}$. This range
represents the velocity spanned by the strongest H~{\footnotesize I} peak,
which is centered at 8114~km~s$^{-1}$.

Figure~4 shows optical-depth images similar to Figure~3, but with
modified gray-scale and contour levels for the weakest H~{\footnotesize I}
components seen in the velocity range 8135.7--8128.8~km~s$^{-1}$ of the spectra
in Figure~2. These features are mainly seen against the weakest parts of the
background continuum, with little optical depth toward the stronger regions.

Our study shows no evidence for absorption against the southern jet or the
weaker northern counterjet of 3C~84, where the flux density is less than 40
and 7~mJy~beam$^{-1}$, respectively.

Figures~5{\it a-e} show images of $N_{\rm HI}/T_{\rm s}$, for the velocity
ranges 8128.3--8125.7, 8124.9--8120.1, 8119.2--8113.1, 8112.7--8110.1, and
8108.8--8104.9~km~s$^{-1}$, respectively. These ranges of velocity correspond
to the six strongest H~{\footnotesize I} absorption features, which are marked
with heavy solid lines in the spectra for regions I, II and VI of Figure~2,
although they can also be seen in some of the other regions. The $N_{\rm
HI}/T_{\rm s}$ images are calculated by integrating over the optical-depth
values in each velocity range, as $N_{\rm HI}/T_{\rm s} = 1.823 \times 10^{18}
\int \tau(v) dv$.

The velocity ranges in Figures~5{\it c} and 5{\it d} are limited so that the 
effect of the blending between the strongest peaks in regions I and II of
Figure~2 is minimized. The velocity range of Figure~5{\it e} covers two
H~{\footnotesize I} absorption peaks that arise at different locations, one in
the east and one in the south (regions II and VI in Figure~2).

Table~2 summarizes the physical characteristics of the six strongest H
{\footnotesize I} absorption features observed against 3C~84.
The velocities (Column~1) refer to peaks of these
features seen in the optical-depth spectra of Figure~2.
The widths of these lines (Column~2) are the approximate full widths at half
maximum optical depth. $N_{\rm HI}/T_{\rm s}$ of each peak (Column~5) is
obtained from the images in Figure~5. The volume densities (Column~7) are 
deduced assuming spherically symmetric clouds with diameters equal to their
observed transverse linear extents, and $T_{\rm s}=50$~K.

\section{DISCUSSION}

The strong background radio source covers an area of about 7 by 16~pc on the
plane of the sky. This area represents a very small section of the high-velocity
foreground galaxy, which has an extent of at least 25~kpc \citep
{RUB77,RUB78,UNG90,VAN77}. However, even in this very compact region, we were
able to identify several H~{\footnotesize I} absorption features.

The nature of the foreground galaxy remains unclear. \citet
{RUB77,RUB78} report an east-west velocity gradient of about 300~km~s$^{-1}$ in
the foreground galaxy, suggesting an edge-on rotating spiral with $V_{\rm
max}=150{\rm~sin}(i)$~km~s$^{-1}$, which is not necessarily in collision with
NGC~1275. However, the observations of \citet {UNG90} show a velocity gradient
in the north-south direction too, by about 100~km~s$^{-1}$, and no evidence for
spiral structure or a single well-defined nucleus. This means that the
foreground galaxy is either an irregular galaxy, as was suggested by \citet
{OOR76}, or a late-type spiral galaxy, disturbed either by its passage through
the intergalactic medium in the core of the cluster, or by its collision with
NGC~1275 \citep {UNG90}. The absence of continuum light from a nucleus or a disk
\citep {VAN77} makes it difficult to determine the exact nature of this
foreground galaxy.

Considering these possibilities for the structure of the intervening
galaxy, we can conduct a comparison of our detected H~{\footnotesize I}
lines with the H~{\footnotesize I} absorption in the Galaxy, as a well
studied spiral, and with H~{\footnotesize I} absorption in the Large Magellanic
Cloud (LMC), as the closest and well known irregular galaxy.

For Galactic H~{\footnotesize I} absorption lines detected against various
background sources, we consider observations obtained with the
Parkes single dish and Parkes two-element interferometer \citep {RAD72,GOS72},
and with the VLA and the Arecibo-Los Ca\~{n}os interferometer \citep {GD89}. The
velocity widths of the Galactic H~{\footnotesize I} absorption lines range
over 1--17~km~s$^{-1}$, the optical depths from 0.11--5.3 and the
derived column densities (0.3--34)$\times 10^{20}$~cm$^{-2}$.
In all three papers, most velocity widths fall in a narrower range, 1--9~km~s$^{-1}$.

In this comparison, we have excluded the VLBA observations of Galactic H
{\footnotesize I} absorption toward background radio sources \citep
{FAI98,FAI01}, where the linear extent of the H~{\footnotesize I} clouds,
between 3 and 100~AU, is much smaller than the extent of the H~{\footnotesize I}
structures in our observations.

The H~{\footnotesize I} absorption lines in the LMC were observed with the
Australia Telescope Compact Array (ATCA) \citep {DIC94,MAR00}. Most of the
H~{\footnotesize I} absorption lines have velocity widths less than
1~km~s$^{-1}$, and very few exceed 1.2~km~s$^{-1}$. The optical-depth values
range between 0.12 and 2.05.

In both galaxies, the Milky Way and the LMC, multi-component H~{\footnotesize
I} absorption features are detected against individual background sources.
The optical depths measured in our extragalactic H~{\footnotesize I}
lines fall in the lower end of the optical depth ranges for both the Galaxy and
LMC\null. While the velocity widths of the Galactic
H~{\footnotesize I} absorption features are close to the values seen in our
observations, which range between 1.5 and 5~km~s$^{-1}$, the features in the LMC
tend to have narrower H~{\footnotesize I} components, with most having widths
less than 1~km~s$^{-1}$.

\section{CONCLUSIONS}

We report the discovery of multiple H~{\footnotesize I} clouds in the
high-velocity ($\sim$8200~km~s$^{-1}$) system associated with NGC~1275 =
Perseus~A, the dominant member of the Perseus cluster (${V_{\rm sys}
\sim}$5200~km~s$^{-1}$).

The background galaxy NGC~1275 is an early-type giant cD elliptical. Its
compact core, known as 3C~84, has a total flux density of $\sim$18~Jy, and a
linear extent of 16~pc on the plane of the sky at 1.38~GHz. Our
observations show a decline of $\sim$4~Jy in the total flux density of 3C~84
since 1995.

The nature of the foreground object remains uncertain. However, all
studies suggest that it is a gas-rich galaxy.
The detected H~{\footnotesize I} lines appear to arise in
``interstellar'' type H~{\footnotesize I} clouds within the foreground galaxy,
and lie along our line of sight to the high-brightness source 3C~84.

The superposition of the foreground galaxy, which lacks a strong radio nucleus,
with the bright core of NGC~1275, provides an unusual opportunity to probe the
properties and physical conditions of extragalactic neutral hydrogen clouds
in a late-type galaxy.
The results show close similarities between these extragalactic H
{\footnotesize I} clouds, and clouds in both the Galaxy and LMC, detected
against various background radio sources.

\section{ACKNOWLEDGMENTS}
The authors thank A.~P.~Sarma and C.~L.~Brogan for their help in the data
reduction. We thank A.~J.~Kemball for helpful discussions on the
template spectrum method in data reduction, and G.~B.~Taylor for guidance on
high dynamic range imaging of 3C~84. We also thank many other members of the
scientific staff of NRAO at Socorro, NM for the helpful ideas while reducing
these global VLBI observations. The European VLBI Network is a joint facility of
European and Chinese radio astronomy institutes funded by their national
research councils. This research has made use of data from the University of
Michigan Radio Astronomy Observatory which is supported by funds from the University of Michigan.
EM is grateful for support from NRAO through the Pre-doctoral Fellowship
Program.

\begin{figure}

\epsscale{0.8}
\plotone{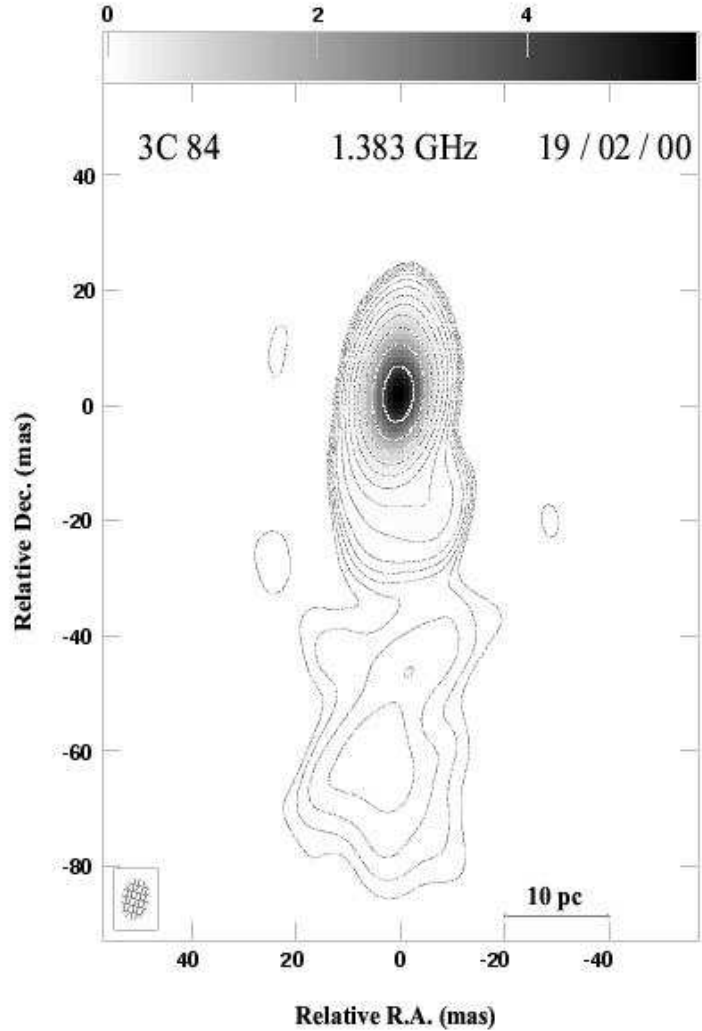}
\caption{Continuum image of 3C~84 at 1383~MHz with its southern jet.
The restoring beam has dimensions $7.2 \times 5.2$~mas in position angle
$-14^{\circ}$. The peak flux is 5.58~Jy~beam$^{-1}$, and the contour levels
are at 4, 5.7, 8, 11, 16, 32, 64,$\ldots$4096~mJy~beam$^{-1}$.
The gray scale range is indicated by the step wedge at the top of the image.
The reference position (0,0) is $\alpha \rm{(J2000)}=03^{\rm h}
19^{\rm m} 48\rlap{.}^{\rm s} 160$, $\delta \rm{(J2000)}=41^{\circ} 30'
42\rlap{.}'' 105$. \label{FIG1}}
\end{figure}

\begin{figure}
\epsscale{0.82}
\plotone{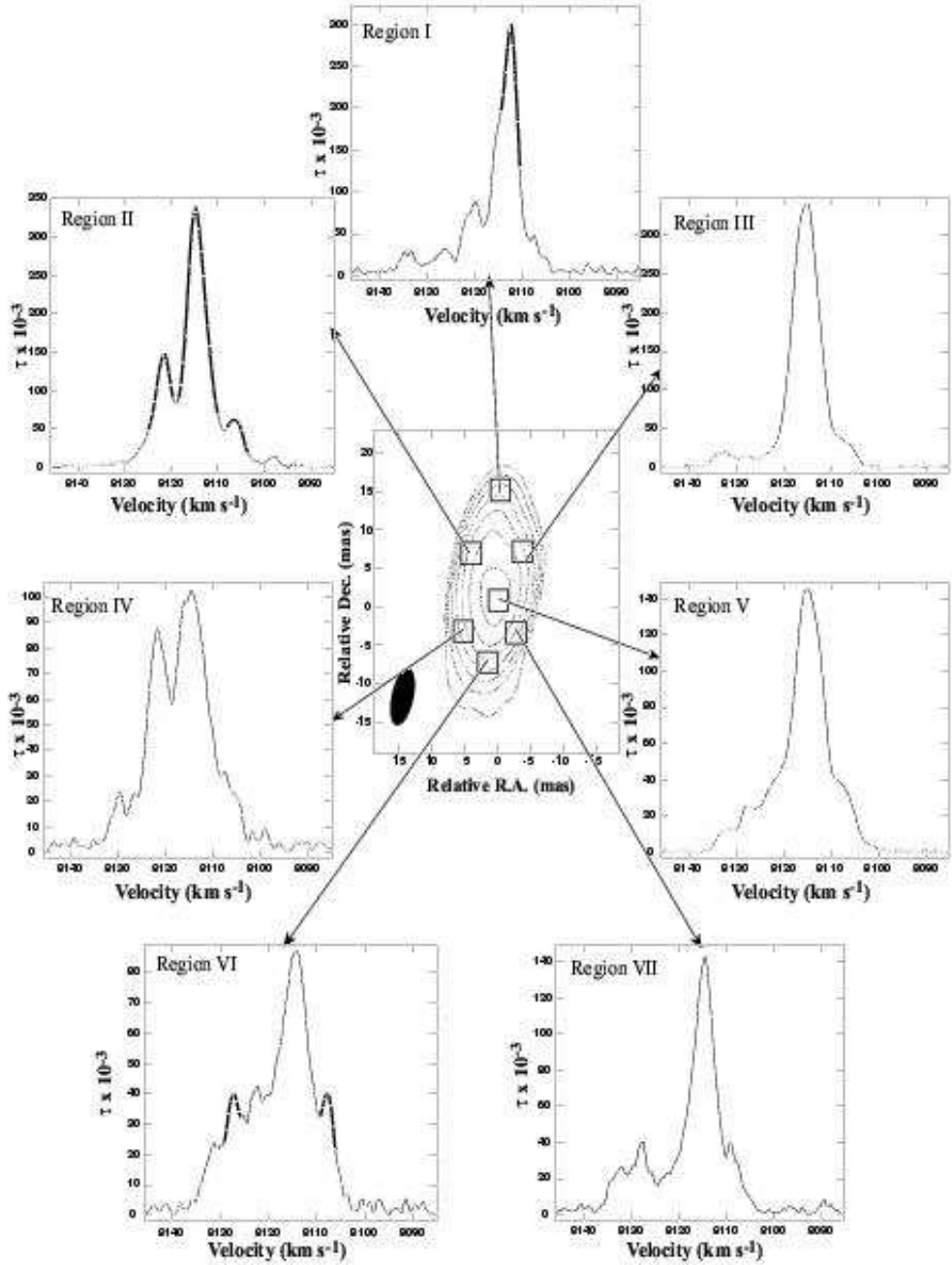}
\caption {H~{\footnotesize I} optical-depth spectra obtained in various
regions against the background continuum source 3C~84 at 1383~MHz. The
restoring beam is $7.66 \times 3.25$~mas in position angle $-14.56^{\circ}$.
The contour levels are at 64, 128, 256,$\ldots$4096~mJy~beam$^{-1}$.
The heavy lines denote the six major absorption components seen toward the
background radio continuum, in the regions where they are most
evident.\label{FIG2}} \end{figure}

\begin{figure}
\epsscale{0.70}
\plotone{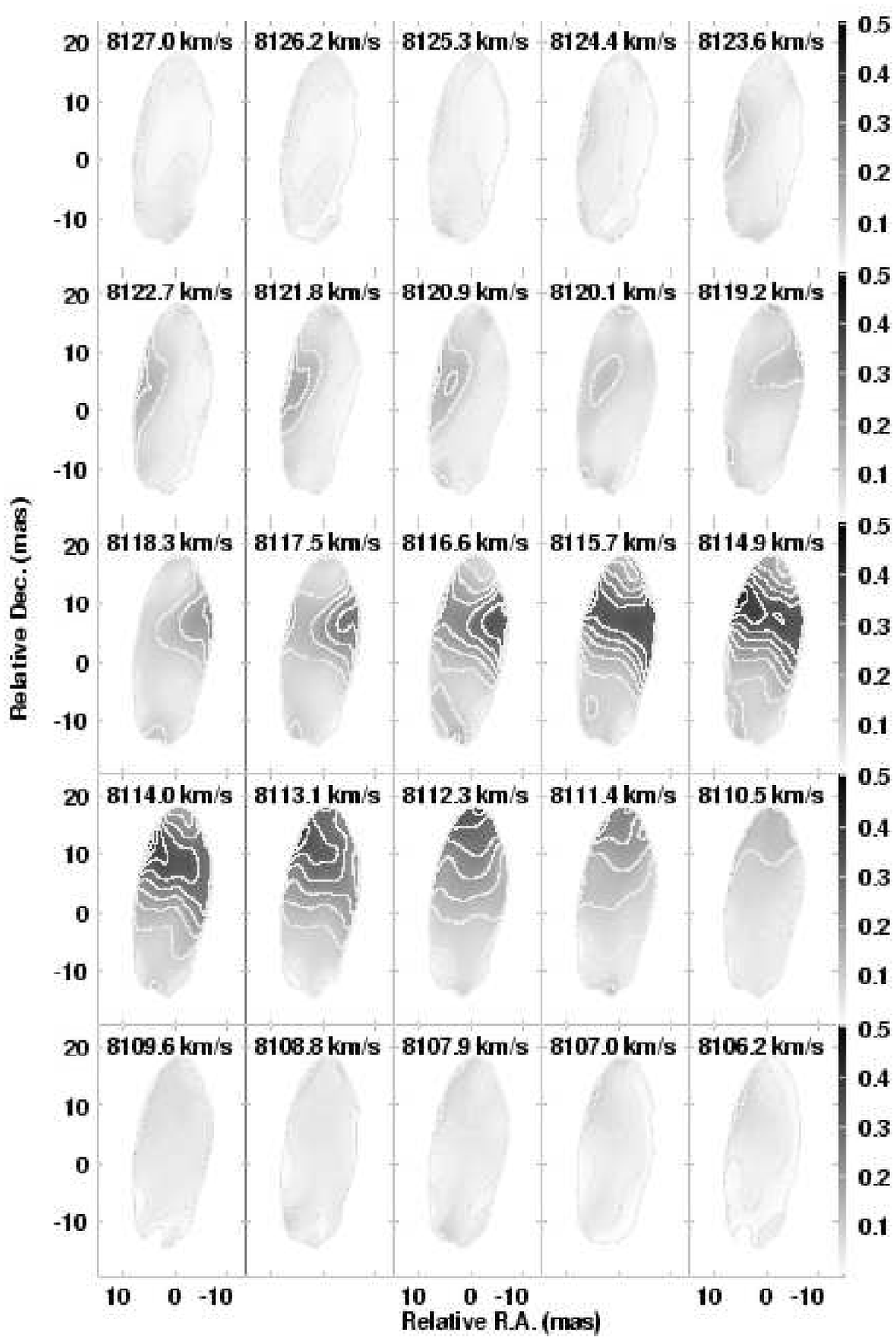}
\caption{Gray-scale and contour H~{\footnotesize I} optical depth channel
images toward 3C~84 in the velocity range 8127.0--8106.2~km~s$^{-1}$, with
every other channel being displayed. The restoring beam in these images is $7.66
\times 3.25$~mas. The gray-scale range is indicated by the step wedge at the
right side of the images; the contour levels are 0.025, 0.5, 0.1,
0.15,$\ldots$0.5.\label{FIG3}} \end{figure}

\begin{figure}
\epsscale{0.6}
\plotone{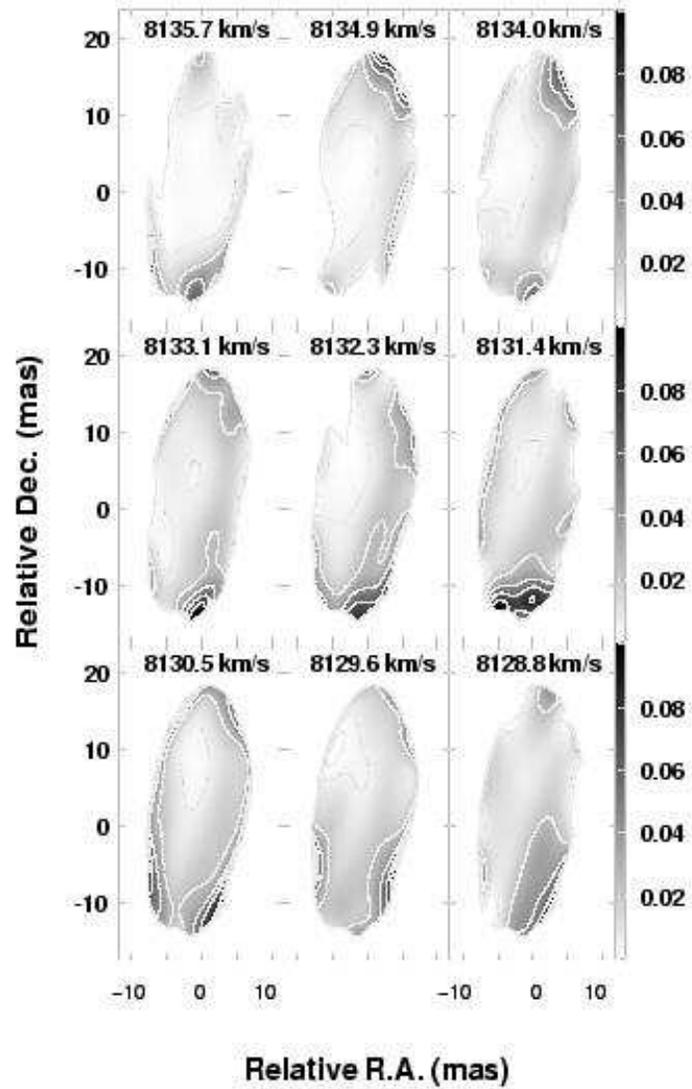}
\caption{Gray-scale and contour H~{\footnotesize I} optical depth channel
images toward 3C~84 in the velocity range 8135.7--8128.8~km~s$^{-1}$, with every
other channel being displayed.  The restoring beam is $7.66 \times 3.25$~mas.
The gray-scale range is indicated by the step wedge at the right side of the
images; the contour levels are 0.005, 0.025, 0.045, 0.065, and 0.085.
\label{FIG4}} \end{figure}

\begin{figure}
\epsscale{0.70}
\plotone{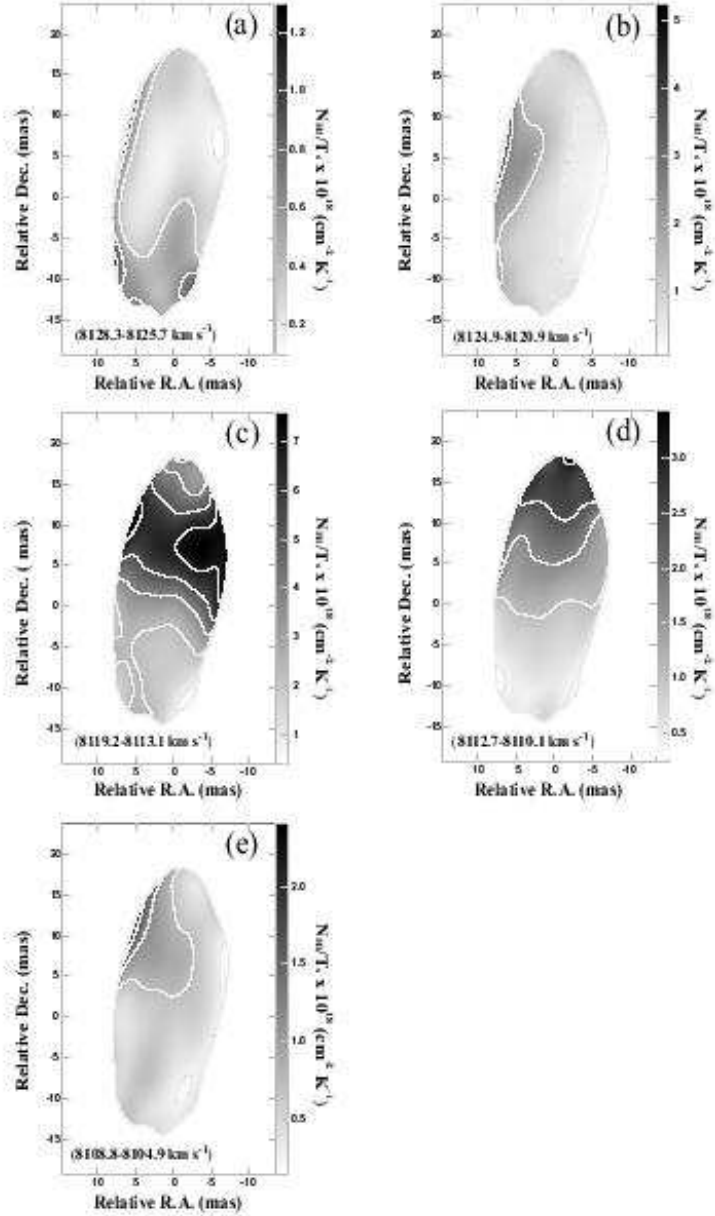}
\caption{Gray-scale and contour $N_{\rm HI}/T_{\rm s}$ images for the velocity
ranges ({\it a}) 8128.3--8125.7~km~s$^{-1}$, ({\it b})
8124.9--8120.1~km~s$^{-1}$, ({\it c}) 8119.2--8113.1~km~s$^{-1}$, ({\it d})
8112.7--8110.1~km~s$^{-1}$, and ({\it e}) 8108.8--8104.9~km~s$^{-1}$. The
contour levels are 10, 30, 50, 70, and 90\% of the peak value indicated by the
step wedge at the right side of each map. The restoring beam in these images is
$7.66 \times 3.25$~mas. \label{FIG5}} \end{figure}

\clearpage

\begin{deluxetable}{lccc}
\tablecolumns{4}
\tablewidth{0pc}
\tablecaption{P{\footnotesize ARAMETERS OF THE} G{\footnotesize LOBAL} VLBI
O{\footnotesize BSERVATIONS}} \tablehead{
\colhead{Parameters} & \colhead{}  & \colhead{} &
\colhead{Values}} \startdata
Observing Date \dotfill & &  & 2000
Feb.~19\\
R.A. (J2000)\dotfill & & &  03 19 48.160 \\
Dec. (J2000)\dotfill & & &  \llap{+ }41 30 42.105 \\

Total observing time (hr)\dotfill &  & & 25 \\

Frequency (MHz)\dotfill & & &  1383 \\

Continuum data bandwidth (MHz)\dotfill  & & & 8 \\

Continuum image rms ($\mu$Jy beam$^{-1}$)\dotfill  & & & 330 \\

Line data bandwidth (MHz)\dotfill  & & & 0.5 \\

Line data velocity coverage (km s$^{-1}$)\dotfill & & &  8058--8169 \\

Line velocity resolution (km s$^{-1}$)\dotfill & & & 0.43 \\

Line image rms (mJy beam$^{-1}$)\dotfill & & & 1.5 \\

Optical depth image cutoff (mJy beam$^{-1}$)\ldots \ldots &  & & 64 \\

\enddata
\tablecomments{Units of right ascension are hours, minutes, and seconds, and
units of declination are degrees, arcminutes, and arcseconds.}
\end{deluxetable}

\clearpage

\begin{deluxetable}{ccclccccc}
\tablecolumns{9}
\tablewidth{0pc}
\tablecaption{P{\footnotesize ARAMETERS OF THE} H~{\footnotesize I}
C{\footnotesize LOUDS}}
\tablehead{
\colhead{Velocity\tablenotemark{a}}   &  \colhead{$\Delta v_{\rm FWHM}$
\tablenotemark{b}} & \colhead{} &  \colhead{$\tau_{\rm max}$} & \colhead{}
& \colhead{Linear Size} & \colhead{$N_{\rm HI}/T_{\rm s}$}
& \colhead{$N_{\rm HI}$\tablenotemark{c}}
& \colhead{$n$\tablenotemark{d}}  \\
\colhead{(km s$^{-1}$)} & \colhead{(km s$^{-1}$)}  & \colhead{} &
\colhead{} &  \colhead{} & \colhead{(pc)} &
\colhead{(cm$^{-2}$ K$^{-1}$)} & \colhead{(cm$^{-2}$)} &
\colhead{(cm$^{-3}$)} \\
\colhead{(1)} & \colhead{(2)} & \colhead{} & \colhead{(3)} &
\colhead{} & \colhead{(4)} & \colhead{(5)} & \colhead{(6)} &
\colhead{(7)}}

\startdata
8126.2 & 1.5  &  &0.14&  & 15  & 1.3 $\times 10^{18}$ & 6.5 $\times 10^{19}$
& \phantom{} 1.4 \\
8121.8 & 3.5  &  & 0.30 &  & 13  & 5.2 $\times 10^{18}$
& 2.6 $\times 10^{20}$ &\phantom{} 6.5 \\
8114.0 & 5.0  &  & 0.45 &  & 10  & 7.6 $\times 10^{18}$& 3.8 $\times
10^{20}$ & 11.2 \\
8112.2 & 3.5  &  & 0.38 & &\phantom{} 9  & 3.4 $\times 10^{18}$ & 1.7
$\times 10^{20}$  &\phantom{} 6.1\\
8108.3 & 1.5  &  & 0.10 & &\phantom{} 5  &  0.7 $\times 10^{18}$ &
3.5 $\times 10^{19}$  &\phantom{} 2.3\\
8106.6 & 3.5  &  & 0.15  &  &\phantom{} 6
& 2.4 $\times 10^{18}$ & 1.2 $\times 10^{20}$  &\phantom{}  6.5 \\
\enddata

\tablenotetext{a}{Heliocentric velocity of the H~{\footnotesize I}
absorption feature at $\tau_{\rm max}$.}
\tablenotetext{b}{Full width at half optical depth.}
\tablenotetext{c}{Based on $T_{\rm s}=50~\rm {K}$.}
\tablenotetext{d}{Assumes spherically symmetric clouds and $T_{\rm s}=50~\rm
{K}$.}
\end{deluxetable}

\end{document}